\documentstyle[twoside]{article}
\def\Red  {}
\def\Black{}
\def\Blue {}

\newcommand{\GeV}{\,{\rm GeV}}
\newcommand{\TeV}{\,{\rm TeV}}

\newcommand{\journal}[4]{{\em #1 \bf #2} (#3) #4}
\newcommand{\NP}[3] {\journal{Nucl. Phys.}{#1}{#2}{#3}}
\newcommand{\PRL}[3]{\journal{Phys. Rev. Lett.}{#1}{#2}{#3}}
\newcommand{\PL}[3] {\journal{Phys. Lett.}{#1}{#2}{#3}}
\newcommand{\PR}[3] {\journal{Phys. Rev.}{#1}{#2}{#3}}
\newcommand{\mhu}{m_{h^{\rm u}}}

\newcommand{\mtR}{m_{\tilde{t}_R}}
\newcommand{\mtL}{m_{\tilde{t}_L}}
\newcommand{\mQt}{m_{\tilde{Q}_3}}
\newcommand{\eq}[1]{~(\ref{eq:#1})}

\def\Ord{{\cal O}}  \def\SU{{\rm SU}}
\def\SO{{\rm SO}} 
\def\circa#1{\,\raise.3ex\hbox{$#1$\kern-.75em\lower1ex\hbox{$\sim$}}\,}
\makeatletter
\newcounter{alphaequation}[equation]
\def\thealphaequation{\theequation\hbox to
0.6em{\hfil\alph{alphaequation}\hfil}}
\def\eqnsystem#1{
\def\@eqnnum{{\rm (\thealphaequation)}}
\def\@@eqncr{\let\@tempa\relax \ifcase\@eqcnt \def\@tempa{& & &} \or
  \def\@tempa{& &}\or \def\@tempa{&}\fi\@tempa
  \if@eqnsw\@eqnnum\refstepcounter{alphaequation}\fi
\global\@eqnswtrue\global\@eqcnt=0\cr}
\refstepcounter{equation} \let\@currentlabel\theequation \def\@tempb{#1}
\ifx\@tempb\empty\else\label{#1}\fi
\refstepcounter{alphaequation}
\let\@currentlabel\thealphaequation
\global\@eqnswtrue\global\@eqcnt=0 \tabskip\@centering\let\\=\@eqncr
$$\halign to \displaywidth\bgroup \@eqnsel\hskip\@centering
$\displaystyle\tabskip\z@{##}$&\global\@eqcnt\@ne
\hskip2\arraycolsep\hfil${##}$\hfil& \global\@eqcnt\tw@\hskip2\arraycolsep
$\displaystyle\tabskip\z@{##}$\hfil
\tabskip\@centering&\llap{##}\tabskip\z@\cr}
\def\endeqnsystem{\@@eqncr\egroup$$\global\@ignoretrue} \makeatother

\begin{document}
\twocolumn[
\centerline{Sept.~1996 \hfill  \bf hep-ph/9609286\hfill
 \bf FT--UAM 96/40} \vspace{1cm}
\centerline{\LARGE\bf\Red Radiatively induced light right-handed stop}

\bigskip\bigskip\Black
\centerline{\large\bf Alessandro Strumia} \vspace{0.3cm}

\centerline{\em Departamento de F\'{\i}sica Te\'orica,
Universidad Aut\'onoma de Madrid,}
\centerline{\em 28049, Madrid, Espa\~na \rm and \em
INFN, sezione di Pisa,  I-56126 Pisa, Italia}

\bigskip\bigskip\Blue

\centerline{\large\bf Abstract}
\begin{quote}\large\indent
A right-handed
stop not much heavier or even lighter than the $Z$ boson
has today desirable phenomenological consequences.
We study how it can result
within the usual radiative scenario of electroweak symmetry breaking.
A restriction on the gaugino mass parameters, $M_2\circa{<}0.3 m_{10}$,
arises if soft terms satisfy relations suggested
by unification theories.
Moreover, requiring to get a light stop without unnatural
fine-tunings below the per-cent level, we obtain
another more interesting upper bound on the
chargino mass, $M_\chi\circa{<}M_Z$,
and we derive interesting conclusions about the masses
of the gluino, $M_3\sim (150\div300)\GeV$,
of the heavy stop, $M_{\tilde{T}}\sim(250\div 500)\GeV$,
and of the left/right mixing angle in the stop sector,
$|\theta_{\tilde{t}}|\circa{<}0.3$.
\end{quote}\Black
\vspace{1cm}]

\noindent
\paragraph{1}
There are today independent phenomenological indications
that suggest the presence of a light, mostly right-handed, stop state
with mass around the $Z$-pole.

To begin, it has recently pointed out in~\cite{B}, that
the electroweak phase transition can be
sufficiently strongly first order 
(so that the observed
baryon asymmetry can be generated during the
electroweak phase transition)
for acceptable values of the higgs mass
if a stop state is light,
or, more exactly, if its soft mass term
is sufficiently small, $|\mtR^2|\sim M_Z^2$.
%% so that $M_{\tilde{t}}\circa{<} 150\GeV$.
%More negative values of $\mtR^2$ give rise to
%dangerous unphysical minima~\cite{B,S[A<3]}.

Moreover, a mostly right-handed stop
with mass $M_{\tilde{t}}\circa{<}100\GeV$,
together with a similarly light chargi\-no,
mediates supersymmetric corrections to the electroweak
precision observables
that explain the discrepancy between the
measured value of the $Z\to b\bar{b}$ width and its Standard
Model (SM) prediction
without affecting the other (successful) SM-predictions~\cite{Rb}.
The consequent decrease of the predicted
hadronic $Z$ width would reduce
the extracted value of the strong coupling constant, $\alpha_3(M_Z)$,
as suggested by its low energy determinations~\cite{Rb,alphas}.

In both cases these desired features are
specific of the small $\tan\beta$ region.
Even if a definitive numerical calculation is still lacking,
with a moderate amount of mixing between
the $\tilde{t}_L$ and $\tilde{t}_R$ states,
the lightest stop $\tilde{t}$ can probably be sufficiently light and
right-handed so that to solve both problems at the same time
without requiring a too large negative value of
the stop soft term $\mtR^2\circa{>} - m_t^2$
that would give rise to charge and color breaking minima
of dangerous kind~\cite{S[A<3]}.

\paragraph{2}
It is well known that since
the higgs doublet that gives mass to up-quarks, $h^{\rm u}$,
is the smallest multiplet
involved in the Yukawa coupling of
the top quark, its soft mass term, $\mhu^2$,
is easily driven to be negative
by renormalization effects.
For universal soft terms at the unification scale
this happens in all the parameter space.
We will study in which regions of the parameter
space this happens also for the soft mass$^2$ term, $\mtR^2$,
of the second smallest
multiplet involved in the top-quark Yukawa coupling,
the right-handed stop.
We will see that requiring to get a light stop
%$\mtR^2\sim -M_Z^2$
in the ordinary successful scenario for the radiative
breaking of the electroweak gauge symmetry,
implies, among the other things, the presence of a very light
chargino in the spectrum and gives strongly favored
values for the gluino mass,
for the heavy stop mass and for the
left/right mixing angle in the stop sector.

The most important new feature regards naturalness bounds~\cite{FT}:
since we now want to fix {\em two\/} different combinations
of soft terms to be around at the $Z$ scale,
the presence of heavy supersymmetric
particles (around at $1\TeV$) in the spectrum
requires now strongly unnatural fine-tuning of the various
relevant parameters.

We only assume that the
gaugino masses satisfy GUT-like relations
and that the soft terms are pointlike operators
up to the unification scale, as happens, for example,
if they are mediated
by supergravity couplings~\cite{SuGraSoft}\footnote{We
will see that in the opposite case~\cite{GaugeSoft}
it is difficult to obtain a light right-handed stop.}.
In this case,
near the infrared fixed point of the top quark Yukawa coupling
and for moderate values of $\tan\beta$,
the $Z$-boson mass is obtained,
from the MSSM minimization conditions,
as a sum of different terms
\begin{equation}\label{eq:MZ^2}
M_Z^2 \approx -2\mu^2 +\Ord(1) m_0^2 + \Ord(10) M_2^2
\end{equation}
where $M_2$ is the $\SU(2)_L$ gaugino mass parameter
and $\mu$ is the `$\mu$ term', both renormalized at the weak scale,
while $m_0^2$ is a typical scalar mass squared at the unification scale.
We will consider unnatural a situation where
the single contributions to $M_Z^2$ in eq.\eq{MZ^2} are much bigger than
their sum, and define the inverse `fine tuning'
$1/f'=\Delta'$ as the largest single contribution to
$M_Z^2$ in units of $M_Z^2$.
In ordinary situations, and near the
infrared fixed-point of the top quark
Yukawa coupling, this simplified definition does not differ
from other reasonable and more sophisticated choices~\cite{FT}
that consider the stability of $M_Z^2$ with respect to variations
of a set of parameters chosen as the `fundamental' ones.
For this reason we will continue to employ the term `fine tuning'
in our discussion of the naturalness bounds.
In particular, since the right handed term of eq.\eq{MZ^2} contains
a $\Ord(10) M_2^2$ term, we have  $\Delta'\circa{>} 10M_2^2/M_Z^2$,
so that, imposing the naturalness condition
$\Delta'<\Delta_{\rm lim}$, implies
$$M_2\circa{<}1\TeV\,(\Delta_{\rm lim}/10^3)^{1/2}.$$

Now we also want to obtain a light right-handed stop.
Since $\mtR^2$ is obtained as
\begin{equation}\label{eq:mtR}
\mtR^2 \approx -\Ord(0.3) m_0^2 + \Ord(6) M_2^2
\end{equation}
we need, in particular, another
$\Delta''\circa{>} 6 M_2^2/M_Z^2$ inverse fine-tuning.

For this reason, imposing a naturalness
bound on the {\em total\/}
`fine tuning', $\Delta=\Delta'\Delta''<\Delta_{\rm lim}$,
gives rise to an upper bound on the sparticle masses
that is stronger and less dependent
on the personal choice of the maximum tolerated `fine tuning',
$1/\Delta_{\rm lim}$.
In particular a significant bound on $M_2$ is obtained:
$$M_2\circa{<}2 M_Z\,(\Delta_{\rm lim}/10^3)^{1/4}.$$
This means that {\em a very light chargino
is consequence of requiring a light stop\/}.
More generally, interesting naturalness upper bounds can be derived
on the parameters, like $M_2$ and $m_0$,
that appear in both the expression
for $M_Z^2$ and for $\mtR^2$.
On the contrary the $\mu$-term ---
and consequently the masses of charged and pseudoscalar Higgs fields --
are less strongly constrained\footnote{
Here we are considering the $\mu$ term as an independent parameter,
even if we expect that in a natural theory
$\mu$ vanishes in the supersymmetric limit.
In unified models with this property,
$\mu$ can naturally be obtained as
a model-dependent linear combination of the dimension-one
soft terms~\cite{muGUT}.}.
A numerical analysis is of course necessary to
clarify the strength of these double `fine-tuning' bounds.

\begin{figure}[t]\setlength{\unitlength}{1cm}
\begin{center}\begin{picture}(7,6.7)
%\put(-0.5,0){\special{picture xy}}
\put(-0.5,0){\includegraphics{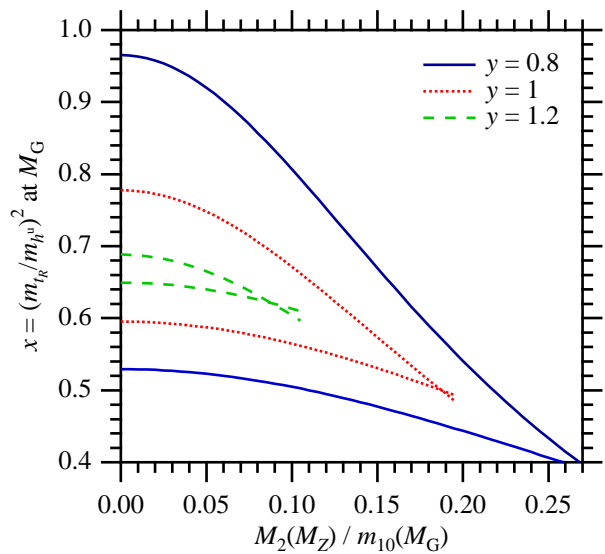}}
\end{picture}
\caption{\em A negative value of the stop soft term $\mtR^2$ at the
Fermi scale
is obtained in the region inside the lines, plotted
for different values of $y\equiv \mtR^2/\mQt^2$ at $M_{\rm GUT}$,
$A_t=0$ at $M_{\rm GUT}$,
and near the infrared fixed point for the top
Yukawa coupling.}
\end{center}
\end{figure}

\paragraph{3}
Let us now consider these issues in a more
quantitative way.
We parameterize the relevant boundary condition at the
unification scale as
\begin{equation}\label{eq:x}
m_{10}^2 \equiv \mtR^2 =y \mQt^2 = x \mhu^2
\end{equation}
where $x$ and $y$ are unknown numbers.
Even if we will keep them arbitrary,
we reasonably expect that $y\approx 1$ in a unified model.
In fact, unless the chiral families mix with extra vector-like states
at the unification scale in such a way that
$\tilde{Q}_3$ is not unified with $\tilde{t}_R$,
the equality between their GUT-scale soft masses,
$\mQt$ and $\mtR$,
can only be broken by (small) GUT threshold corrections.
%This requires the interplay between $M$ and $v$,
%and if the GUT scale is dynamically induced
%$M\sim v$ can be not natural.
%Also FCNC problems...
On the contrary there is no reason for imposing $x=1$, because,
even if were justifiable at the Planck scale, this choice
would be spoiled by (unknown) GUT renormalization effects
and, in SO(10) models, also by the $D$-term
of the broken $\SO(10)/\SU(5) = {\rm U}(1)_X$ factor.

\begin{figure}[t]\setlength{\unitlength}{1cm}
\begin{center}\begin{picture}(7,6.7)
%\put(-0.5,0){\special{picture FTch}}
\put(-0.5,0){\includegraphics{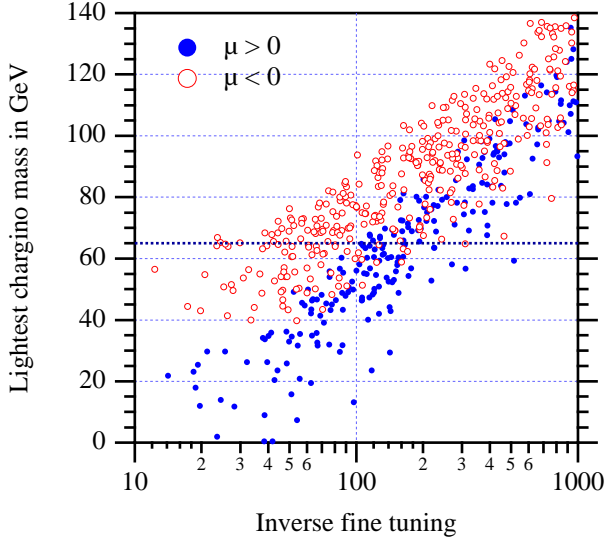}}
\end{picture}
\caption{\em Fine tuning as function of the chargino mass.
The horizontal line is the present experimental lower bound
on the chargino mass.}
\end{center}
\end{figure}

\medskip

The parameters at the electroweak scale are obtained as~\cite{RGE}
\begin{eqnsystem}{sys:RGE}
\mhu^2 &=& \mhu^2(M_{\rm GUT})+ 0.79 M_2^2-{\textstyle 1\over2}3I\\
\mQt^2 &=& \mQt^2(M_{\rm GUT})+ 10.6  M_2^2-{\textstyle 1\over6}3I\\
\mtR^2 &=& \mtR^2(M_{\rm GUT})+ 10.0  M_2^2-{\textstyle 1\over3}3I
\end{eqnsystem}
with
\begin{eqnarray*}
3I  &=&  \rho [X_t(M_{\rm GUT})+3\rho (6.5-2.5\rho)M_2^2]+\\
&&+\rho (1-\rho) [A_{t\rm G}^2 +5.5 M_2 A_{t\rm G}],
\end{eqnarray*}
where $X_t\equiv \mQt^2+\mtR^2+\mhu^2$,
$A_{t\rm G}$ is the trilinear term for the top Yukawa coupling
at the unification scale,
and
$\rho=(\lambda_t(M_Z)/\lambda_{tZ}^{\rm max})^2$,
where $\lambda_{tZ}^{\rm max}\approx 1.14$ is the
maximum value that $\lambda_t(M_Z)$
can reach without developing a Landau pole below
the unification scale.
Note that $\rho\approx 1$ near
the infrared fixed point for $\lambda_t$,
so that the dependence of the scalar masses
on $A_{t\rm G}$ is negligible.
Note also that the gaugino contribution to the soft masses
does not depend
on the unknown boundary condition for the scalar masses.
This shows that the fine-tuning constraints on $M_2$
are not model-dependent.

Requiring that the $\tilde{t}_R$ and $h^{\rm u}$ soft mass$^2$ terms
be negative give rise to the restrictions on the parameter space
shown in figure~1.
Let us explain its features in a qualitative way.
Above the upper lines (at big $x$) $\mtR^2$ is so big
that cannot be driven to be negative by the $\lambda_t$
effect in the renormalization group equations (RGEs).
Below the lower lines (at small $x$) the same happens for $\mhu^2$,
preventing the breaking of the electroweak gauge group.
As shown in fig.~1, for moderate values of $y$ the two bounds are
compatible only in a restricted range
of $x$, $0.5\circa{<}x\circa{<}0.9$,
and for small values of $M_2\circa{<}m_{10}/4$, where
$M_2$ is renormalized at the electroweak scale.
This particular restriction on
the parameter space also suggests the presence of
a light chargino.
Note also that with universal soft terms
it is not possible to get a sufficiently light right-handed stop,
at least unless $A_{t\rm G}$ is very large.

\medskip

We now plot, in figure~2, the values of the lightest chargino mass
as function of the required fine-tuning, as defined previously,
for arbitrary random possible values of the various parameters
$\lambda_t$, $\tan\beta$, $\mu$, $m_{10}$, $M_2$, $A_{t\rm G}$,
$x$ and with $y$ in the range $0.5\div 1.5$.
We see that a lightest chargino not very light,
$M_\chi\circa{>}M_Z$, is possible only with
a considerable amount of fine tuning,
beyond the \%{} level and that it is impossible to require a fine tuning
weaker than the $10\%$ level, as done in~\cite{FT},
without contradicting the experimental lower bound on the chargino mass.
Since the lightest chargino is very light,
finite one loop quantum corrections that we have not included
can increase its mass by $(3\div10)\%$~\cite{Mchi1loop}.

\smallskip

We now discuss the stop sector, finding a nice consistency
of the various bounds.
The solution of the RGE for $A_t$ is~\cite{RGE}
$$A_t = (1-\rho) A_{t\rm G} + (4.9-2.8\rho) M_2$$
so that $A_t\sim (1\div3.5) M_2$
near the infrared fixed point for
the top Yukawa coupling\footnote{The stronger forms for the $M_2/A_t$
and $M_2/\mu$ correlations
that appear in the literature~\cite{rad} hold
for large values of $M_2$ that, in our case, are
disfavored by naturalness bounds.}.
This is a welcome restriction.
In fact $A_t$ cannot be too large,
both because the condition
$$A_t^2+3\mu^2 \circa{<} 7.5 (\mtR^2+\mtL^2)$$
must be satisfied to avoid a too fast decay of
the SM-like minimum into charge and
color breaking minima~\cite{S[A<3]},
and both in order that the lightest eigenstate
of the stop mass matrix
$$\pmatrix{\mQt^2 + m_t^2 + 0.35 M_Z^2 & -m_t(A_t^*+\mu\cot\beta) \cr
-m_t(A_t+\mu^*\cot\beta) & \mtR^2+m_t^2 + 0.15 M_Z^2\cr}$$
be almost right-handed (for $\mtR^2 \circa{<} 0 < \mQt^2$).
We now plot in figure~3, again
for random samples of the free parameters and near the
infrared fixed point of the top quark Yukawa coupling,
the possible values of the gluino mass and of
the heaviest stop mass,
distinguishing the cases of
moderate fine tuning, $\Delta<100$, (big green
dots), strong fine tuning, $100<\Delta<1000$, (medium orange
dots) and very strong fine tuning, $\Delta>1000$, (small red
dots).
We have imposed the experimental lower bounds on
the masses of the supersymmetric particles, and
in particular the bound on the chargino mass,
$M_\chi>65\GeV$.
Since $A_t$ is small and positive
(due to its correlation with $M_2$), while
$\mu$ has most likely
the opposite sign (see fig.~2), and for
the values of the heaviest stop mass suggested in figure~3,
the left/right mixing angle $\theta_{\tilde{t}}$ in the stop sector
is small, $|\theta_{\tilde{t}}|\circa{<}0.3$.
It is interesting that this is the appropriate amount of mixing
necessary to get a sufficiently light and right-handed stop state,
$m_{\tilde{t}}\circa{<}100\GeV$,
without encountering problems
with dangerous unphysical minima~\cite{B,S[A<3]}
or with undesired radiative corrections~\cite{Rb}.
In conclusion,
we summarize the naturalness bounds on
$M_3$ and $M_{\tilde{T}}$ as
$$M_3\circa{<}300 \GeV \sqrt[4]{\frac{\Delta_{\rm lim}}{100}},\qquad
M_{\tilde{T}}\circa{<}500 \GeV \sqrt[4]{\frac{\Delta_{\rm lim}}{100}}.$$
Finally, the dependence of $M_Z^2$
and $\mtR^2$ on the masses of the sfermions of
first and second generation
only comes, at one loop,
through a small term, neglected in eq.~(\ref{sys:RGE}),
proportional to $X_Y\equiv \sum N_R Y_R m_R^2$,
where the sum runs over all the MSSM scalar fields $R$
with $N_R$ components and hypercharge $Y_R$.
The associated fine-tuning
upper bound,
$X_Y\!\circa{<}\! \TeV (10/\Delta_{\rm lim})^{1/2}$,
has been computed in ref.~\cite{12FT}.
Adding the requirement of a light stop it becomes
$X_Y\circa{<} \TeV (100/\Delta_{\rm lim})^{1/4}$.

\begin{figure}[t]\setlength{\unitlength}{1cm}
\begin{center}\begin{picture}(7,6.7)
%\put(-0.5,0){\special{picture M3stop}}
\put(-0.5,0){\includegraphics{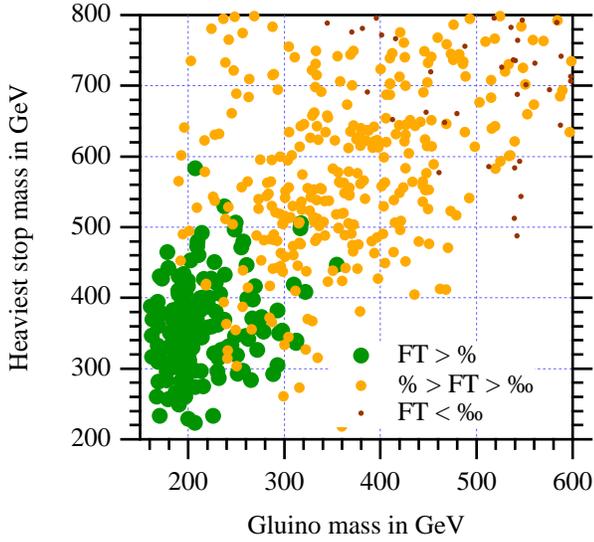}}
\end{picture}
\caption{\em Values of $(M_3,M_{\tilde{T}})$ for
weak, strong or very strong fine-tuning {\rm FT}.}
\end{center}
\end{figure}

\paragraph{4}
It is also possible that,
differently from what we have assumed so far,
the soft terms appear
as pointlike terms only up to some energy $M_U$
below the unification scale.
For example supersymmetry breaking could be directly felt by
appropriate gauge charged `transmitter' fields with mass $\sim M_U$ below
or around the gauge-unification scale~\cite{GaugeSoft}.
If this is the case unification physics would
not be reflected in the soft terms,
and minimal models of this kind agree on the predictions
\begin{equation}
M_i\propto\alpha_i,\qquad
m_R^2 \propto c^R_i \alpha_i^2\qquad \hbox{at }M_U
\end{equation}
where $c^R_i$ are the traces, in the representation $R$,
of the generators of the three factors $G_i$ of the SM gauge group.
In particular we have $\mQt^2\sim\mtR^2\sim2\mhu^2$ at $M_U$, so that
we can see from fig.~1 that
it seems not possible to get a light right-handed stop in this case,
neither if $M_U$ were not much smaller than $M_{\rm GUT}$.
We must however remind that in such models it is difficult to
understand the origin of the supersymmetric $\mu$-term,
so that some important amount of non-minimality
(for example in the messenger field content~\cite{GaugeSoft2}
or in the light fields~\cite{NMSSM})
could play a decisive role.

\end{document}